\documentclass[aip,amsfonts,amsmath,amssymb,reprint]{revtex4-1}
\usepackage{graphicx}
\usepackage{dcolumn}
\usepackage{bm}

\usepackage{xcolor}
\usepackage[english]{babel}
\usepackage[utf8]{inputenc}
\usepackage{tabularx,booktabs}

\begin{document}

\preprint{---}

\title[]{Hysteresis and criticality in hybrid percolation transitions}
\author{Jinha Park}
\affiliation{CCSS, CTP and  Department of Physics and Astronomy, Seoul National University, Seoul 08826, Korea}
\author{Sudo Yi}
\affiliation{School of Physics, Korea Institute for Advanced Study, Seoul 02455, Korea}
\author{B. Kahng}
\email{bkahng@snu.ac.kr}
\affiliation{CCSS, CTP and  Department of Physics and Astronomy, Seoul National University, Seoul 08826, Korea}

\date{\today}

\begin{abstract}
Phase transitions (PTs) are generally classified into second-order and first-order transitions, each exhibiting different intrinsic properties. For instance, a first-order transition exhibits latent heat and hysteresis when a control parameter is increased and then decreased across a transition point, whereas a second-order transition does not. Recently, hybrid percolation transitions (HPTs) are issued in diverse complex systems, in which the features of first-order and second-order PTs occur at the same transition point. Thus, the question whether hysteresis appears in an HPT arises. Herein, we investigate this fundamental question with a so-called restricted Erd\H{o}s--R\'enyi random network model, in which a cluster fragmentation process is additionally proposed. The hysteresis curve of the order parameter was obtained. Depending on when the reverse process is initiated, the shapes of hysteresis curves change, and the critical behavior of the HPT is conserved throughout the forward and reverse processes. 
\end{abstract}

\maketitle
\begin{quotation}
A hybrid phase transition (discontinuous, but with criticality), occurs in several complex systems. In such systems, the question arises whether hysteresis occurs. However, for hybrid synchronization transition hysteresis is absent. Herein, we demonstrate that a hysteresis arises with critical behavior in a hybrid percolation transition by using the Erd\H{o}d--R\'enyi model with global suppression. The critical behavior is related to the associated one in the forward process.
\end{quotation}

Phase transitions are generally classified into second- and first-order transitions~\cite{jaeger}. A second-order phase transition is a continuous transition that is characterized by a critical point at which correlations over all scales hold the system to a unique critical phase. A first-order phase transition is a discontinuous transition that allows two phases to coexist at the transition point. In addition, correlations are finite in their ranges and critical phenomena are absent. At the transition point, the dynamic transition from one phase to another may occur through an equilibrium or nonequilibrium process. Depending on the type of process and the driving rate of the control parameter, a first-order transition may exhibit a single jump corresponding to the Maxwell construction, or exhibit hysteresis behavior~\cite{control_rate}.
In contrast, superheated/supercooled solutions are totally absent in some discontinuous equilibrium phase transitions. A notable example is the landscape inversion phase transition of colloidal metamagnets~\cite{landscape_inversion,pnas_magnet}. The free energy landscape becomes flat at (and inverted across) the transition point. In such a case, the metastable branch of solutions are absent, i.e., a hysteresis curve does not exist and the phase transition is hybrid. 

A hybrid phase transition is a discontinuous transition but with criticality. Such a transition occurs in several complex systems, including cascade failure of interdependent networks~\cite{havlin}, $k$-core~\cite{kcore,kcore_mendes,kcore_lee}, cooperative epidemic spreading~\cite{contagion1,contagion2,contagion3,golden} and synchronization~\cite{pazo,pcKM,acKM,basnarkov,Song}. However, a possible hysteresis together with critical behavior in hybrid phase transitions has been rarely reported thus far, even though hysteresis in discontinuous transitions was investigated in explosive synchronization~\cite{kurths14,boccaletti14,boccaletti16} and percolation~\cite{invasion_percolation,bootstrap_dhar,reverse_bootstrap,explosive_panos,mixed_hysteresis,nanotube,kim,spanning,sumrule_bastas} in complex systems.     

For synchronization, the Kuramoto model with uniform natural frequency distribution exhibits a hybrid phase transition~\cite{pazo}. A frequency-entrained giant cluster suddenly emerges at the transition point by an abrupt frequency locking. In turn, the order parameter jumps and exhibits a non-integer exponent $\beta=2/3$~\cite{pazo,pcKM,acKM,basnarkov}. Thus, the transition is hybrid. A recent study showed that an $ad$-$hoc$ potential~\cite{Song}, proposed analogously to the Landau free energy, exhibits a flat landscape as it appears in colloidal metamagnetic systems~\cite{landscape_inversion,pnas_magnet}. The order parameter is uniquely determined except at the critical point. Hence, hysteresis is absent. The $ad$-$hoc$ potential approach may be useful to predict whether or not a hybrid phase transition occurs for nonlinear dynamic systems.

For percolation, hybrid percolation transitions (HPTs) may be classified into two categories: one occurring during the cascade process as in $k$-core percolation ~\cite{havlin,kcore,kcore_mendes,kcore_lee,contagion1,contagion2,contagion3,golden}, and the other during the cluster merging process~\cite{r-er,cho,IET}. For the former case, dynamics proceed only in one direction of decreasing occupation probability, and a reverse process may not be well defined, i.e., a cascade cluster aggregation process is hardly imagined. Thus, it is hardly be checked if hysteresis exists. For the latter case, however, a reverse process is assigned from a cluster fragmentation process~\cite{fraction}. Thus, the existence of a hysteresis may be checked. In this paper, we consider a reverse process of the cluster merging dynamics of the so-called restricted Erd\H{o}s--R\'enyi ($r$--ER) model. It is known that the $r$--ER model~\cite{r-er,cho,IET} exhibits an HPT. In the reverse fragmentation process, a restriction is given to the breakage of large clusters. Hence, both forward and reverse $r$--ER processes are similar in that either the growth or collapse of large clusters is being suppressed.

The cluster coagulation process of the $r$--ER model is dichotomous, depending on the set classification ($A$ and $B$). At each time, clusters are ranked by their sizes and classified into two sets $A$ and $B$, which contain a portion $gN$ ($0< g\leq 1$) of nodes of the smallest clusters and the remaining large clusters, respectively. Two nodes are selected for connection: one node is chosen randomly from set $A$, and the other is chosen randomly from set $A$ or $B$. They are then connected by a link unless they are already connected. Subsequently, the classification is updated as the cluster rankings are changed. This cluster merging process is repeated. The time variable is defined as $t=L/N$, where $L$ is the number of occupied links and $N$ is the number of nodes in the systems. This $r$--ER model includes two important factors: i) the growth of large clusters is practically suppressed, because a node belonging to small clusters has twice the chance to be linked, whereas a node in large clusters has a single chance; and ii) the dynamic rule becomes global in the process of sorting out the portion of the smallest clusters among all cluster sizes. Owing to these two factors, an HPT occurs at $t_c$. 

To characterize the HPT of the $r$--ER model in finite systems, two characteristic times $t_g$ and $t_c$ were introduced for finite systems~\cite{IET}: $t_g$ is the time at which the fraction of nodes in a giant cluster first exceeds the capacity $1-g$ of set $B$. At and beyond $t_g$, the partition is removed and the two sets are unified. Further, at $t_c$ ($t_c > t_g$), the size distribution of finite clusters exhibits power-law decay, $n_s\sim s^{-\tau}$. The exponent $\tau$ is continuously varying on the control parameter $g$. The power-law behavior of the cluster size distribution at $t_c$ in the forward process indicates the critical behavior of HPT. The two times $t_g$ and $t_c$ are not the same in finite systems, but reduce to the same point in the limit $N\to \infty$.

The reverse process is implemented as follows: 
\begin{enumerate}
\item[(i)] As a preparatory step, the $r$--ER model is processed up to a given return time $t_r 
\ge t_g$, and then the reverse process is launched. 
\item[(ii)] At each time ($t\equiv L/N$), clusters are ranked by their sizes and classified into two sets $A$ and $B$, which contain a portion $gN (0< g\leq 1)$ of nodes of the smallest clusters and the remaining large clusters, respectively. 
\item[(iii)] A link is randomly chosen. If the removal of this link does not divide a cluster into two smaller clusters that remain both in set $B$ after the rank-based re-classification of clusters, proceed to the removal of selected link. Otherwise, go back to (ii).
\item[(iv)] If a link is removed, the time is decreased by $1/N$. Repeat the processes (ii) and (iii) until the system reaches an absorbing state.
\end{enumerate}
Another possible implementation is replacing the above step (iii) as follows:
\begin{enumerate}
\item[(iii$^\prime$)] A link is randomly chosen. If the removal of this link does not divide a cluster in set $B$ into two smaller clusters that will both move to set $A$ after the re-classification, proceed to the removal of selected link.
\end{enumerate}
In the use of (iii), the fragmentation cannot proceed any further after reaching the absorbing state of step (iv), in which all clusters in set $A$ are of unit size. In contrast, if (iii$^\prime$) is used, the reverse process can proceed down to $t=0$.
The suppression process (iii) or (iii$^\prime$) is symbolically represented as $c_B\not\rightarrow c_B+c_B$ or $c_B\not\rightarrow c_A+c_A$, respectively. Consequently, both $c_B\rightarrow c_A+c_B$ and $c_B\rightarrow c_A+c_A$ are allowed for (iii), and both $c_B\rightarrow c_A+c_B$ and $c_B\rightarrow c_B+c_B$ are allowed for (iii$^\prime$). Further, for both cases, $c_A\rightarrow c_A+c_A$, $c_A\to c_A$, and $c_B\to c_B$ are allowed.

\begin{figure}[!thb]
	\centering\includegraphics[width=\linewidth]{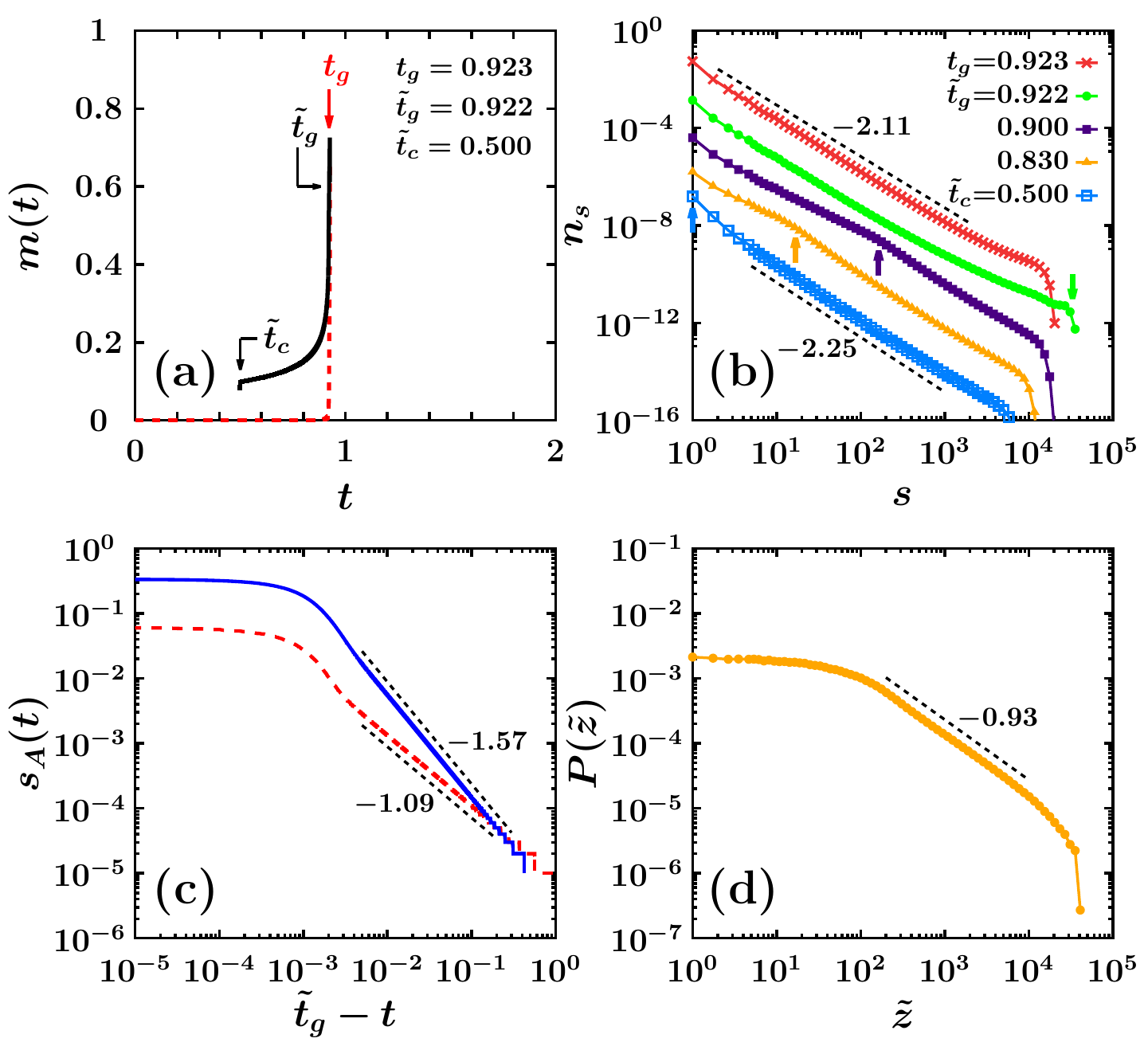} 
	\caption{(a) Hysteresis of the order parameter $m(t)$ through the forward (dashed curve) and backward (solid curve) processes of the $r$--ER model for the case $c_B \not \to c_B+c_B$ and the return time taken as $t_r=t_g$. (b) The cluster-size distribution $n_s$ versus $s$. As the reverse process is proceeded, the bump on $n_s$ moves toward $s=1$. Arrows indicate $s_A(t)$ for given times, representing the size of the largest cluster in set $A$. For visualization, the data of different times are shifted vertically. (c) $s_A$ versus $\tilde{t}_g-t$ (solid curve) for the reverse process, which exhibits power-law decay $s_A(t)\sim (\tilde{t}_g-t)^{-1/\tilde{\sigma}}$ with exponent $1/\tilde{\sigma} \approx 1.57$. $s_A$ versus $t_g-t$ (dashed curve) for the forward process, which exhibits power-law decay $s_A(t)\sim (\tilde{t}_g-t)^{-1/\sigma}$ with exponent $1/\sigma \approx 1.09$. (d) The inter-event time distribution $P(\tilde z)\sim \tilde z^{-\tilde{\alpha}}$ with $\tilde \alpha \approx 0.93$.  Simulations were performed for $g=0.4$ and $N=10^5$ and averaged over $N_{samp}=10^5$ samples.}
	\label{fig1}
\end{figure}

For the forward process, the growth of the giant is suppressed, so that the transition point is delayed and an HPT occurs. For the reverse process, the property of cluster fragmentation depends on the rule (iii). For case (iii), there exists an absorbing state, in which set $A$ is filled with clusters of size one. The time at which such a state is created is denoted as $\tilde{t}_c$ (hereafter, the notation ``tilde'' indicates a reverse process). When the absorbing state occurs, set $A$ cannot accommodate any new size one clusters generated by the process $c_B\rightarrow c_A+c_B$ and $B$ cannot accommodate either owing to the suppressive rule $c_B\not\rightarrow c_B+c_B$. Thus, the fragmentation cannot take place and the dynamics are frozen. This absorbing state can be reached while the order parameter remains finite. Interestingly, a critical behavior appears near $\tilde{t}_c$ when $t_r\approx t_g$, but not when $t_r \gg t_c$. However, for case (iii$^\prime$), the absorbing state is absent. Here, $c_B\rightarrow c_B+c_B$ is allowed; hence, fragmented clusters of smaller sizes can be accommodated in set $B$, if not in set $A$. Thus, a large cluster can be divided into smaller clusters without falling into the absorbing state and the order parameter can be reduced to zero. For each case, hysteresis curves of the order parameter are shown in Fig.~\ref{fig1}(a). At the starting point $t=t_r$ of the reverse process, all clusters including the giant cluster are regarded as the components of set $A$, because the size of the giant cluster exceeds the capacity of $B$, i.e., $(1-g)N$. As the giant cluster gradually breaks down into smaller clusters, it fits into the capacity of set $B$, for which time is denoted as $\tilde{t}_g$.   

\begin{figure}[!htb]
	\centering\includegraphics[width=\linewidth]{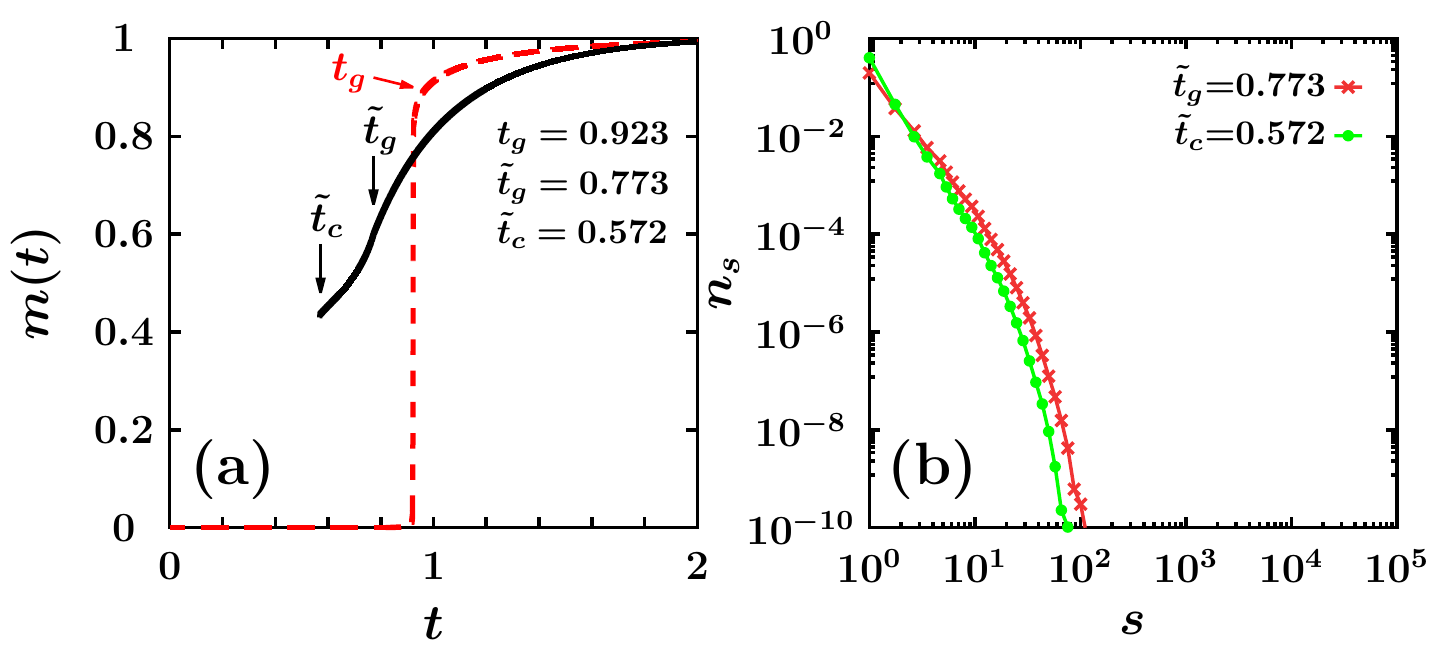} 
	\caption{(a) Hysteresis of the order parameter $m(t)$ through the forward (dashed curve) and reverse (solid curve) processes of the $r$--ER model for the case $c_B \not \to c_B+c_B$ and the return time taken as $t_r \gg t_g$. (b) The cluster-size distribution $n_s$ versus $s$. $n_s$ exhibits exponential decay behavior. Simulations were performed for $g=0.4$ and $N=10^5$ and averaged over $N_{samp}\approx 10^4$ samples.}
	\label{fig2}
\end{figure}
\begin{figure}[!htb]
	\centering\includegraphics[width=\linewidth]{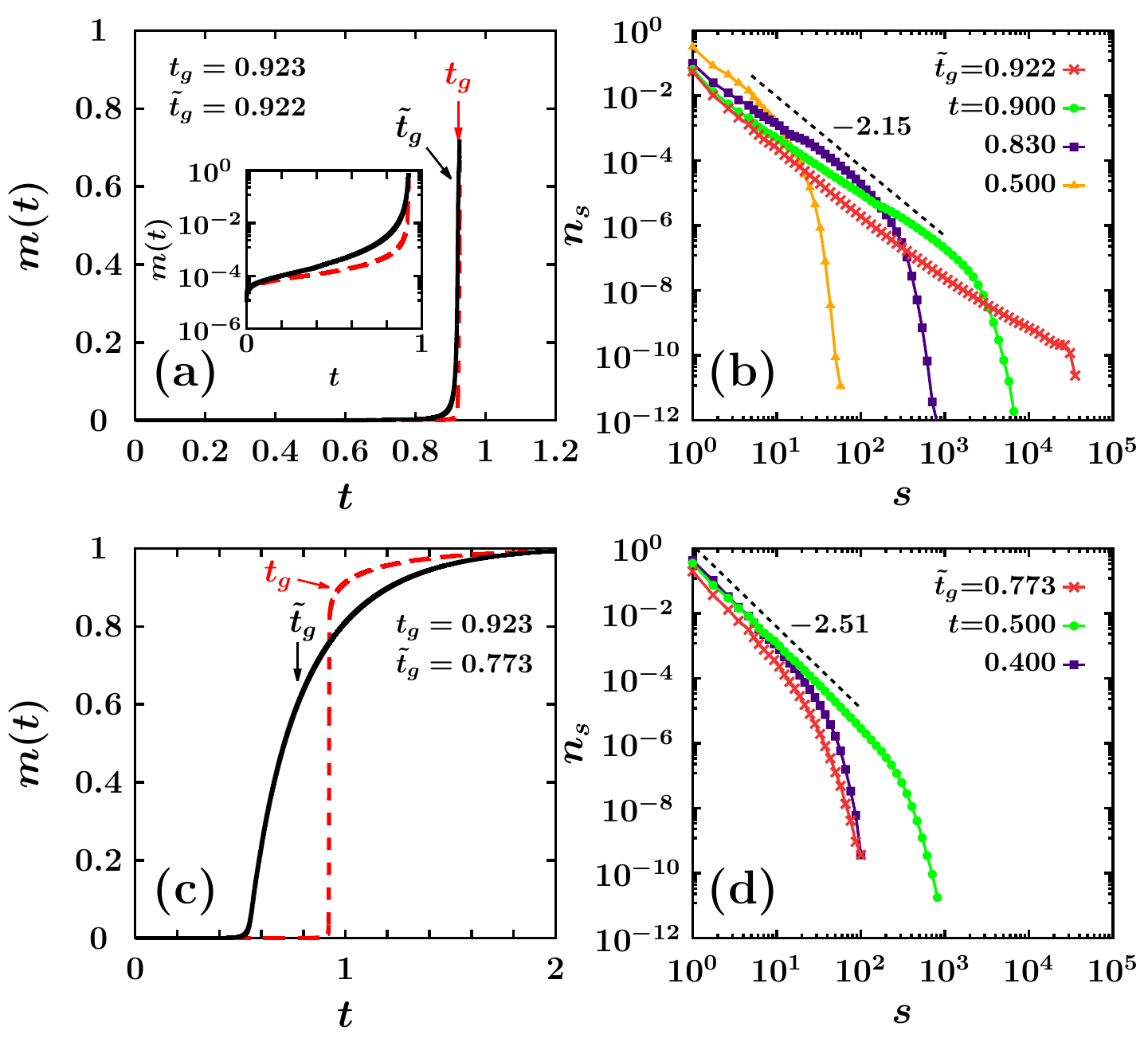} 
	\caption{Hysteresis of the order parameter $m(t)$ through the forward (dashed curve) and backward (solid curve) processes of the $r$--ER model for the case $c_B \not \to c_A+c_A$ and the return time taken as (a) $t_r=t_g$,  and (c) $t_r \gg t_g$. For (a), hysteresis, although small, is noticed. A semi-log plot was used for clarity. The cluster-size distribution $n_s$ versus $s$: (b) immediately after the launch of reverse process at $t_g$, $n_s$ exhibits fat-tailed behavior at $\tilde{t}_g$. Shortly thereafter, it changes to power-law decay, with the exponent $\tau$ of the forward process, followed by exponential cut-off. The power-law region becomes shorter in time because of the reverse process; and (d) $n_s$ exhibits power-law decay behavior with exponential cut-off. The slope is estimated as $\approx -2.5$, consistent with the ER value. Simulations were performed for $g=0.4$ and $N=10^5$ and averaged over $N_{samp}\approx 10^4$ samples.}
\label{fig3}
\end{figure}

Herein, we investigate underlying mechanism of cluster fragmentation microscopically. We first consider case (iii). We consider the following two aspects separately: (1) when $t_r=t_g$, and (2) when $t_r \gg t_g$. \\
When $t_r=t_g$, the forward process is stopped at $t_g$, and the reverse process is launched. The order parameter decreases suddenly; however, the cluster size distribution $n_s$ exhibited power-law decay in a small cluster-size region, but contains a bump in the tail part. Thus, the system had not yet gone through the second-order transition, which is a part of the hybrid phase transition. As soon as the reverse process is launched, $\tilde{t}_g$ is reached and clusters are separated into two sets $A$ and $B$ containing small and large clusters, respectively. Here, $n_s$ is fat-tailed and has a bump at the end, as shown in Fig.~\ref{fig1}(b). The probability $p_\ell$ of choosing a link in the giant cluster is large, and the process $c_B\to c_A+c_B$ mainly occurs. Thus, the giant cluster is divided into a small cluster and a remaining large cluster, and its size is therefore reduced. As such processes are repeated, the bump at the fat-tail of $n_s$ shifts towards a smaller-cluster-size position, owing to the process $c_B\to c_A+c_B$, and the remaining tail part developed a power-law. The bump position in $n_s$ is close to the size of the largest cluster in set $A$, denoted as $s_A(t)$ and marked by arrows in Fig.~\ref{fig1}(b). Fig.~\ref{fig1}(c) shows that this characteristic size scales as $s_A(t)\sim (\tilde{t}_g-t)^{-1/\tilde{\sigma}}$. This feature also occurred in the forward $r$--ER process~\cite{IET}. When the reverse process reaches the absorbing state at $\tilde{t}_c$, the bump disappears, and $n_s(\tilde{t}_c)\sim s^{-\tilde{\tau}}\exp(-s/s^*)$, where $s^*$ is cut-off by the finite-size effect. The exponents $\tilde \tau$ and $\tilde{\sigma}$ are a new pair of critical exponents for the reverse process. As shown in Fig.~\ref{fig1}(c), $\tilde \sigma$ is slightly smaller than $\sigma$, implying that the disintegration process near $\tilde{t}_c$ goes on more gradually than the merging process near $t_c$, which is consistent with the slow-down behavior of the order parameter as shown in~Fig.~\ref{fig1}(a). Further, $\tilde{\tau}$ is slightly larger than $\tau$, because $c_B\to c_A+c_B$ fragments large clusters into smaller ones.

Moreover, we consider the inter-event time distribution composed of the intervals of inter-set crossing events $A\leftrightarrow B$ for each node. It exhibits power-law decay $P(\tilde{z})\sim \tilde{z}^{-\tilde{\alpha}}$. As shown in Figs.~\ref{fig1}(b), (c), and (d), the exponent $\tilde{\alpha}$ satisfies the scaling relation $\tilde{\alpha}=4-\tilde{\tau}-\tilde{\sigma}$ with some small numerical errors, analogous to the dynamic scaling relation $\alpha=4-\tau-\sigma$ for the forward process, which was rooted in the power law of the selection probability of clusters~\cite{IET}.

\begin{table*}[!htb]
  \renewcommand{\arraystretch}{1.2}
  \centering 
  \newcolumntype{C}{>{\centering\arraybackslash}X}
  \begin{tabularx}{\textwidth}{@{}*{11}{C}@{}}
    \hline\hline
    $g$ & $\tilde{t}_c$ & $ \tilde{t}_g$ & $t_g$ & $\tilde{\tau}$ & $\tilde{\sigma}$ & $\tilde{\alpha}$ & $\tau_*$ & $\tau$ & $\sigma$ & $\alpha$ \\
    \hline
    0.2 & 0.728 & 0.983 & 0.984 & 2.16 & 0.63 & 0.98 & 2.13 & 2.08 & 0.97 & 1.03 \\
    0.3 & 0.610 & 0.956 & 0.957 & 2.22 & 0.64 & 0.95 & 2.18 & 2.12 & 0.94 & 1.05 \\
    0.4 & 0.500 & 0.922 & 0.923 & 2.27 & 0.65 & 0.93 & 2.23 & 2.16 & 0.91 & 1.07 \\
    0.5 & 0.398 & 0.882 & 0.882 & 2.31 & 0.65 & 0.91 & 2.26 & 2.18 & 0.89 & 1.08 \\ 
    0.6 & 0.302 & 0.836 & 0.837 & 2.35 & 0.64 & 0.89 & 2.26 & 2.18 & 0.87 & 1.09 \\ 
    0.7 & 0.213 & 0.784 & 0.785 & 2.37 & 0.63 & 0.88 & 2.26 & 2.18 & 0.84 & 1.09 \\ 
    0.8 & 0.132 & 0.723 & 0.725 & 2.37 & 0.63 & 0.86 & 2.35 & 2.25 & 0.81 & 1.09 \\
    \hline\hline 
  \end{tabularx}
  \caption{Table of exponent values: $\tilde \tau$ was obtained at $t=\tilde{t}_c$ in the reverse $r$--ER process under the rule $c_B \not \to c_B+c_B$. Return time is taken as $t_r=t_g$. Further, $\tilde \sigma$ and $\tilde \alpha$ are the exponents of the reverse process. The exponents $\tau_*, \tau, \sigma$, and $\alpha$ are for the forward process, which are also listed for comparison. $\tau_*$ and $\tau$ were obtained at $t=t_c$ by solving the self-consistency equation and by simulations, respectively~\cite{cho}. $\sigma$ and $\alpha$. $\tilde{t}_c$, $\tilde{t}_g$, $t_g$, and all exponents but $\tau_*$ were estimated in finite systems of size $N=10^5$.}
  \label{table1}
\end{table*}

We found that the exponents for the reverse process are continuously varying, as summarized in Table~\ref{table1}. $\tilde{\tau}(g)$ increases with $g$, which is consistent with the $g$-dependent behavior of the forward $\tau(g)$. In contrast, $\tilde{\sigma}(g)$ is related to the dynamics near the absorbing state, and hence it is nearly constant. As shown in Fig.~\ref{fig1}(b), the bump quickly shifts to the small-cluster regime, after which the self-organized critical dynamics near the absorbing state of Fig.~\ref{fig1}(c) follow.

When $t_r \gg t_g$, the giant cluster becomes more densely connected at the return time than at $t_g$, where it was tree-like. As a result, the giant cluster is more resistant to fragmentation, and as shown in Fig.~\ref{fig2}(a), the order parameter decays more gradually near the return time $t_r\gg t_g$ compared to the return at $t_r=t_g$, where the order parameter drops fast from the beginning of the reverse process in Fig.~\ref{fig1}(a). Until $\tilde{t}_g$, the order parameter in the reverse process decays in a similar manner to that of the original ER model. When the system reaches the time $\tilde{t}_g$, the partition between the two sets $A$ and $B$ isi recovered. However, notably, $\tilde{t}_g$ is neither close to the critical point $t_c=0.5$ of the ER model nor the forward transition point $t_g$. Consequently, the size distribution of finite clusters at $t_g$ follows that of the ER model in the supercritical regime $n_s(\tilde{t}_g)\sim s^{-\tilde \tau}\exp(-s/s^*)$, where $s^*$ is rather small because $\tilde{t}_g$ is far away from the transition point of the ER model, as shown in Fig.~\ref{fig2}(b). Beyond $\tilde{t}_g$, the fragmentation dynamics differs from that of the ER model because of the presence of the partition; however, the order parameter decreases continuously until $\tilde{t}_c$. The characteristic cluster size $s^*$ remains small for the late time returns $t_r \gg t_g$. Thus, the criticality near $\tilde{t}_c$ does not appear. 

For case (iii$^\prime$), when $c_B\not\to c_A+c_A$, the absorbing state is absent. Here, the set $A$ with full occupation of size one clusters is not an absorbing state, because any clusters in $B$ of size greater than unity may keep on dividing by $c_B\to c_B+c_B$. Accordingly, the dynamics kept processing beyond $\tilde{t}_c$ towards $t=0$. The order parameter decreases along the trajectory followed by the ER pruning process. Thus, the order parameter decreases continuously to zero, and a hysteresis exists and the hysteresis loop has an ``8" shape. The cluster-size distribution exhibits power-law decay with exponent $\tilde{\tau}=5/2$ at a transition point of the ER model as the criticality of the ER model. All these properties are shown in Fig.~\ref{fig3}.

\begin{figure}[!htb]
	\centering\includegraphics[width=\linewidth]{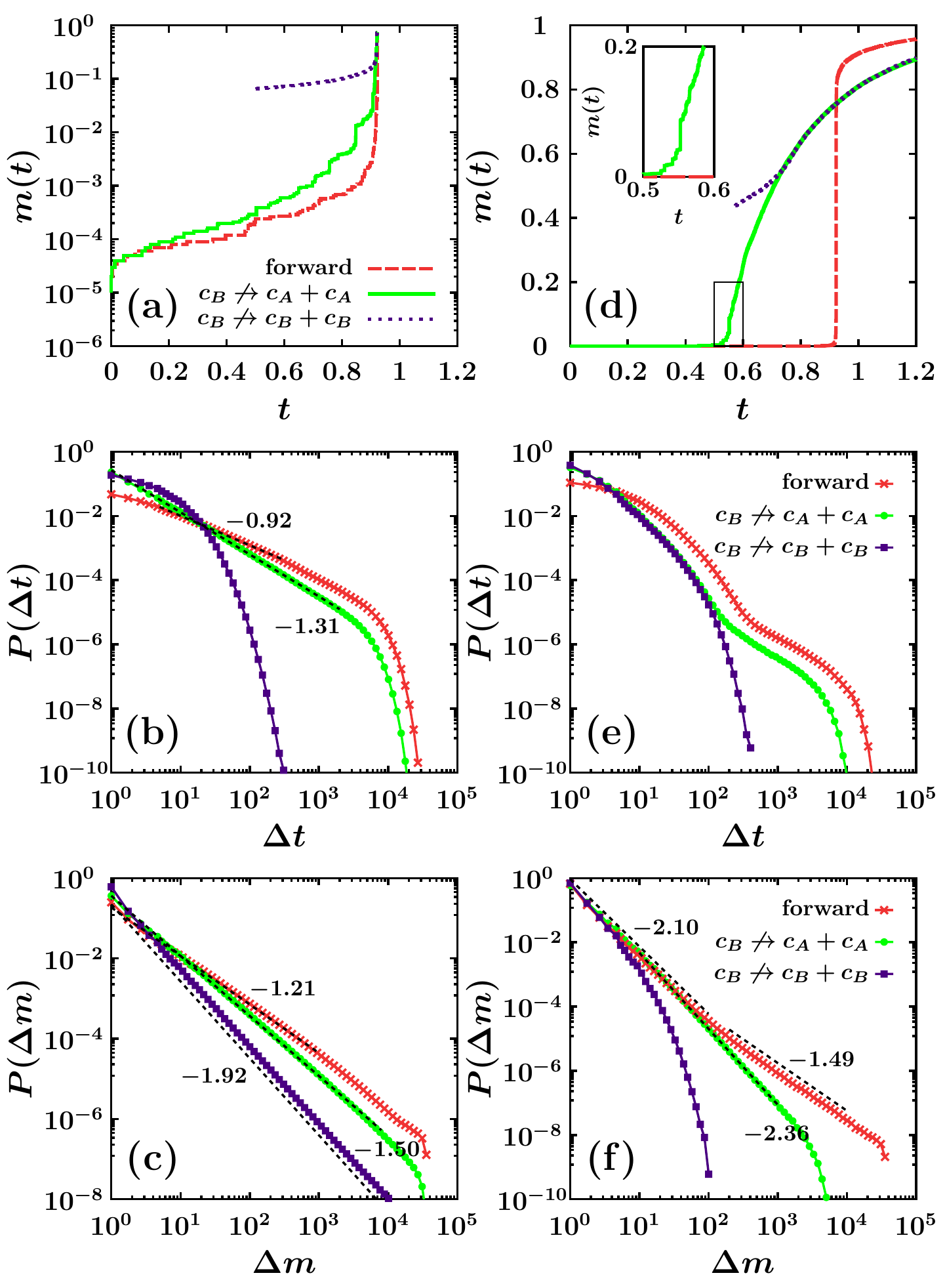} 
	\caption{Devil's staircase of the largest cluster during the fragmentation processes for a single configuration: (a) hysteresis of the order parameter $m(t)$ for a single configuration in the forward (red, bottom dashed curve) and reverse processes of the $r$--ER model for the cases $c_B \not \to c_B+c_B$ (indigo, top dotted curve) and $c_B \not \to c_A+c_A$ (green, middle solid curve) when the return was launched at $t_r=t_g$. The distributions of (b) the widths (time differences) and (c) the height differences of the devil's staircase. Simulations were performed for $g=0.4$ and $N=10^5$. The distributions (b) and (c) were averaged over $N_{samp}\approx 10^5$ samples. (d)--(f) are the correspondences to (a)--(c) for when the reverse process was launched at $t_r=2 > t_c$, and were averaged over $N_{samp}\approx 10^4$ samples.}
	\label{fig:fig4}
\end{figure} 

We traced the largest-cluster size during the reverse process in the system. For a single configuration, it decreases discontinuously, similar to a staircase with irregular steps. The height differences $\{\Delta m_i \}$ and widths $\{\Delta t_i \}$ of the staircase were measured for all the cases of reverse process. The distributions of $\{\Delta m_i \}$ and $\{\Delta t_i \}$ exhibit power-law decay or Poisson-type behavior. The details are presented in Fig.~\ref{fig:fig4}. When the order parameter decreases gradually (drastically), the exponent of the power-law decay is larger (smaller) than two.

In conclusion, the formation of a hysteresis of the order parameter for a hybrid percolation transition induced by cluster merging process using the $r$--ER model was investigated. We found that although the hysteresis exists, the decreasing pattern of order parameter changes significantly depending on the reverse process. The reverse process of the restricted cluster merging model can be categorized into two types: an encouragement and a discouragement of the fragmentation of large clusters. For the former case, the following observations were made: \\
 1) When the forward process was stopped at $t_g$, so that the order parameter exhibited a first-order transition but the criticality was not formed yet, and the reverse process was launched, the order parameter decreased rapidly, which then gradually approached an absorbing state. Thus, a hysteresis was generated. Due to the dichotomous rule of the $r$--ER model, a critical behavior appeared near the absorbing transition point. This critical behavior originated from the tug-of-war process between the two partitioned sets of small and large clusters. Thus, the scaling relation between the inter-event time distribution and the cluster-size distribution, which was established in the forward process, was also satisfied in the reverse process. \\
 2) When the forward process ran far beyond a transition point and was stopped, so that the order parameter experienced both the first-order and second-order transitions, and the reverse process was then launched, the order parameter decreased gradually and entered into an absorbing state without exhibiting any critical behavior. \\ 
On the other hand, for the latter case, i.e., when the fragmentation of large clusters were discouraged, the reverse process became similar to the pruning process of the ER model. Thus, the order parameter decreased continuously and a hysteresis was created. In summary, the hysteresis exists with critical behavior for the $r$--ER model, exhibiting a hybrid percolation transition induced by cluster-merging dynamics. However, the pattern of the hysteresis curve depends on the detailed rules of reverse process. The critical behavior is related to that of the forward process. Therefore, when an undesirable giant cluster is formed abruptly, the reverse process must be turned on immediately to eliminate the giant cluster; otherwise, restoration to the original state takes a significantly longer time.

\begin{acknowledgements}
This work was supported by the National Research Foundation of Korea (NRF) through Grant Nos. NRF-2014R1A3A2069005 (BK). JP and SY equally contributed to the paper as the first author. 
\end{acknowledgements}

\section*{data availability}
The data that support the findings of this study are available from the corresponding author upon reasonable request.

\end{document}